# A 1.5GS/s 8b Pipelined-SAR ADC with Output Level Shifting Settling Technique in 14nm CMOS

Yuanming Zhu, Shengchang Cai, Shiva Kiran, Yang-Hang Fan, Po-Hsuan Chang,
Sebastian Hoyos, and Samuel Palermo
Analog and Mixed-Signal Center
Department of Electrical and Computer Engineering, Texas A&M University
College Station, TX, USA
Email: {zymk1989, spalermo}@tamu.edu

*Abstract*—A single channel 1.5GS/s 8-bit pipelined-SAR ADC utilizes a novel output level shifting (OLS) settling technique to reduce the power and enable low-voltage operation of the dynamic residue amplifier. The ADC consists of a 4-bit first stage and a 5-bit second stage, with 1-bit redundancy to relax the offset, gain, and settling requirements of the first stage. Employing the OLS technique allows for an inter-stage gain of ~4 from the dynamic residue amplifier with a settling time that is only 28% of a conventional CML amplifier. The ADC's conversion speed is further improved with the use of parallel comparators in the two asynchronous stages. Fabricated in a 14nm FinFET technology, the ADC occupies 0.0013mm$^2$ core area and operates with a 0.8V supply. 6.6-bit ENOB is achieved at Nyquist while consuming 2.4mW, resulting in an FOM of 16.7fJ/conv.-step.

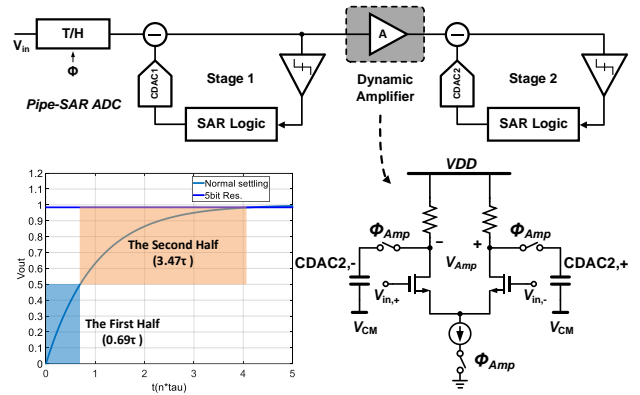

Fig. 1. Pipeline SAR ADC with a dynamic residue amplifier.

## I. INTRODUCTION

As wireline communication data rates climb about 100Gb/s, there is an increased number of receiver front-ends utilizing high-speed analog-to-digital converters (ADCs) that allow for subsequent powerful digital equalization and symbol detection techniques [1][9][10][11][12]. These converters employ a large amount of time-interleaved unit-ADC channels to achieve the required high effective sampling rates. Sampling in these ADCs is typically performed in several ranks with multiple clock phases per rank and generating these clocks with the required accuracy and jitter performance is a major challenge [13]. Another major issue is maintaining the required signal bandwidth as the interleaving factor is increased. This motivates the development of high-speed low-power unit ADCs that can reduce the interleaving factor for a given effective sampling rate, resulting in smaller area and an overall simpler design.

Successive-approximation-register (SAR) ADC architectures are popular due to their low comparator count and simple digital logic content, making them suitable for compact and power-efficient mid-resolution time-interleaved ADCs [2]. However, the conversion speed is limited in the most common implementation of the successive approximation algorithm that performs sequential single-bit conversion cycles. As shown in Fig. 1, introducing pipelining in the SAR ADC provides improved speed by decreasing the number of conversion cycles per input sampling event. A critical block in this architecture is the amplifier that transfers the residue signal between the two pipeline stages. In high-speed converters, conventional opamp-based amplifiers are not suitable due to the excessive static power required to meet settling time requirements. An alternative approach is to use a dynamic residue amplifier that is only activated once over the entire conversion process [3][8].

While dynamic residue amplifiers have the potential to save power, these topologies require a small τ to achieve fast settling times. Satisfying this and maintaining a given gain can result in large dynamic tail current values and increased input transistors that load the first pipeline stage capacitive digital-to-analog converter (CDAC). Given that the smallest CDAC that satisfies the kT/C noise requirement is desired to reduce input buffer power, this loading can cause significant reference attenuation that must be compensated with an increased range reference buffer that is difficult to implement with low supply voltages. Another issue is kickback noise due to the coupling through the large dynamic amplifier input transistors.

This work presents a single channel 1.5GS/s 8-bit pipelined-SAR ADC that utilizes a novel output level shifting (OLS) settling technique to enable low-voltage operation of the dynamic residue amplifier with low hardware overhead. A detailed discussion of the proposed OLS settling technique that allows for an inter-stage gain of ~4 with a settling time that is only 28% of a conventional CML amplifier is given in Section II. Section III provides an overview of the asynchronous ADC architecture and key circuit details. Measurement results from a 14nm CMOS FinFET prototype are presented in Section IV. Finally, Section V provides the conclusion.

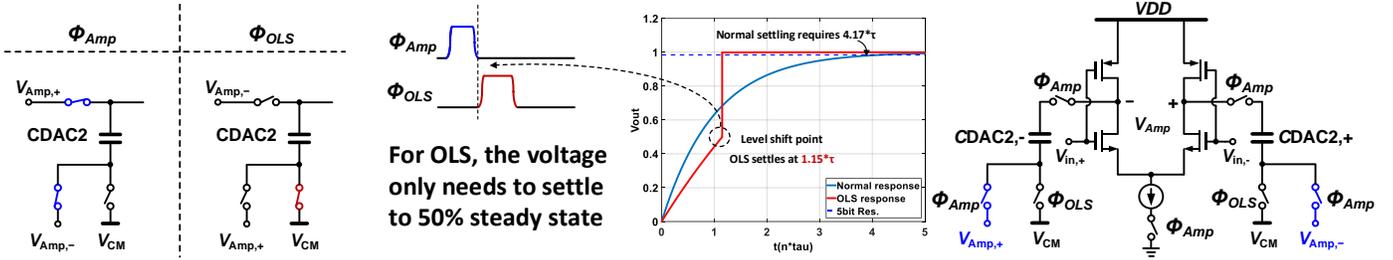

Fig. 2. Output level shifting settling technique and residue amplifier implementation.

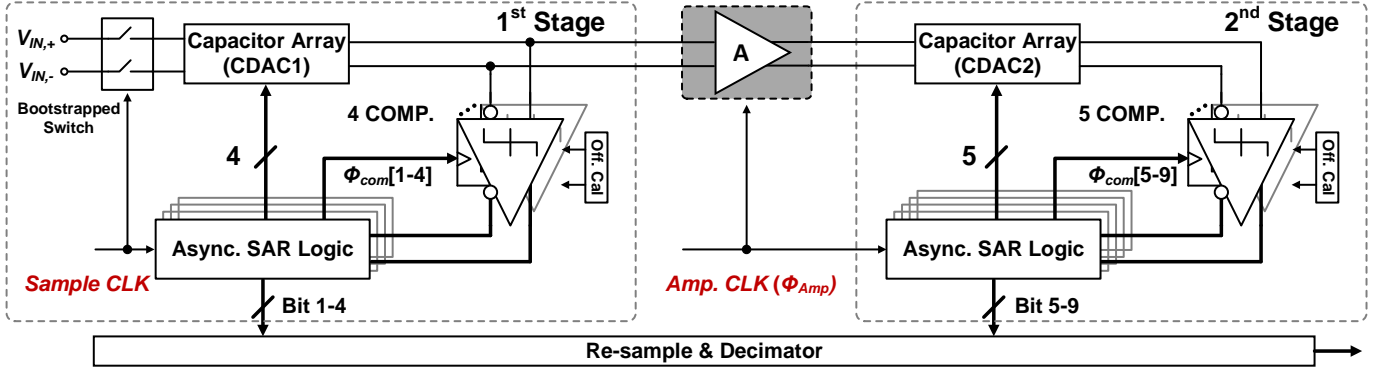

Fig. 3. Pipelined-SAR ADC with output level shifting residue amplifier.

## II. OUTPUT LEVEL SHIFTING SETTLING

While the pipeline-SAR ADC architecture reduces the required settling accuracy of the residue amplifier, it is still challenging to achieve this at high speeds. Upon activation, the Fig. 1 conventional dynamic amplifier output will settle as $V_{Amp} = A_{CML}V_{in}\left(1 - e^{-\frac{t}{\tau}}\right)$, where $A_{CML}$ is the gain. This settles to 50% of the steady state value in a rapid $0.69\tau$, but requires an additional $3.47\tau$ to settle to the 5-bit accuracy required in the second pipeline stage. The brute force method of reducing this settling time is to reduce the load resistor to decrease $\tau$, but this leads to the aforementioned issues of increased tail current values and large input transistor sizes.

Previously, a OLS technique was developed to reduce errors in feedback amplifiers that occur from finite opamp gain [4]. In that work, an initial estimate of the desired output voltage is sampled on a level shifting capacitor and then this capacitor is switched in series with the opamp output and the feedback amplifier output to improve settling accuracy. This work modifies this technique to dramatically improve the settling time of the open-loop dynamic residue amplifier by utilizing the second pipeline stage CDAC2 as the level shifting capacitor. Fig. 2 gives an overview of the proposed OLS settling technique. When $\Phi_{Amp}$ is high and the amplifier is activated, the differential output voltage is stored on both sides of CDAC2 by connecting the nominal amplifier output to the top plate and the opposite output to the bottom plate. This $\Phi_{Amp}$ duration should nominally match the rapid 50% settling time. After this, $\Phi_{OLS}$ is enabled to switch the CDAC2 bottom plate to the common mode. Charge conservation during this phase produces a rapid doubling of the amplifier output signal. Thus, the amplifier output voltage only needs to initially settle to 50% of the steady-state value and the long second half settling is avoided. The significant speed-up offered by the OLS technique is achieved with the low hardware overhead of only one extra bottom-plate CDAC2 switch.

A simplified schematic of the OLS dynamic amplifier, which offers several improvements relative to a conventional CML dynamic amplifier, is shown in the right side of Fig. 2. Instead of utilizing a simple resistive-loaded differential pair, this inverter-based amplifier structure provides both PMOS and NMOS transconductance to provide a higher gain of $A = (g_{mp} + g_{mn})(r_{on}//r_{op}) \approx 2A_{CML}$ at lower supply voltages. While the amplifier has high impedance outputs, a stable output common mode is achieved by resetting the CDAC2 top plate to the common mode prior to activation. One downside of this OLS amplifier is that the equivalent capacitive loading is 4X larger than the conventional CML amplifier due to both sides of CDAC2 being connected to each amplifier output and each capacitor experiencing Miller multiplication. Considering this, the time for the OLS amplifier to achieve 50% settling relative to the original CML amplifier is

$$Av_{in}\left(1 - e^{-\frac{t}{4*\tau}}\right) = 0.5A_{CML}v_{in}$$

$$t = 1.15\tau.$$

Due to the increased amplifier gain, this is only 28% of the $4.17\tau$ required by the conventional CML amplifier at 5-bit resolution. This also results in lower average power due to the dynamic amplifier's reduced activation time. Moreover, the required amplifier's linear output swing range is decreased by a factor of two.

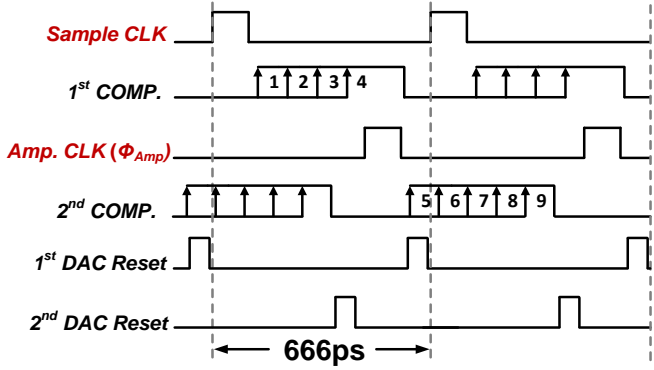

Fig. 4. ADC timing diagram.

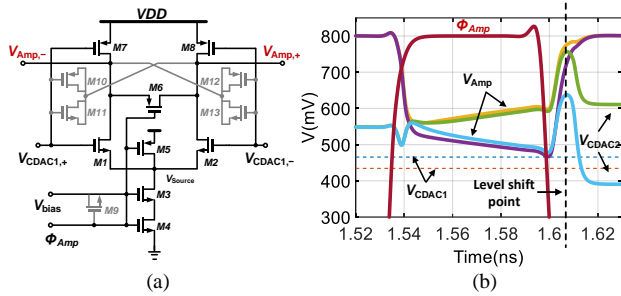

(a) (b)
Fig. 5. Inverter-based dynamic amplifier. (a) Schematic and (b) Simulation results.

One potential issue with the proposed amplifier is matching the duration of $\Phi_{Amp}$ with the 50% settling point. However, high precision is not necessary, as any inaccuracy simply results in a modified gain value that is easily compensated with adjustment of the second stage reference voltages. The jitter specifications on the $\Phi_{Amp}$ signal are also not prohibitive, as a relatively large 8ps$_{pp}$ jitter can be tolerated to achieve 5-bit accuracy at 1.5GS/s.

## III. ADC ARCHITECTURE AND BUILDING BLOCKS

Fig. 3 shows the 8-bit pipelined-SAR ADC with the first pipeline stage converting 4-bits and the second stage converting 5-bits. This 1-bit redundancy between the two stages relaxes the gain, offset, and reference settling requirements of the first stage. The input signal is sampled with a boot-strapped switch that reduces the input sampling time constant and improves high-frequency linearity. kT/C noise requirements are satisfied with CDAC1 and CDAC2 set at 32fF and 16fF, respectively. Both stages employ parallel comparators that are asynchronously activated sequentially for each conversion step, eliminating the comparator reset delay and offering significant speed-up.

The ADC timing diagram is shown in Fig. 4. After input sampling, the first stage converts 4 bits and holds the residue voltage for partial amplification when $\Phi_{Amp}$ goes high. $\Phi_{Amp}$ then transitions low and the level-shifted gain is achieved when the $\Phi_{OLS}$ pulse is activated with minor modifications in the second stage reference switch logic. The second stage then converts the final 5 bits. Both stage CDACs are reset after their conversions are complete to avoid memory effect, with these reset signals internally generated by the comparator ready signal and input clocks. Independent flipped-voltage followers serve as buffers for the two sets of CDAC reference voltages that are locally decoupled with MOS capacitors. As previously mentioned in

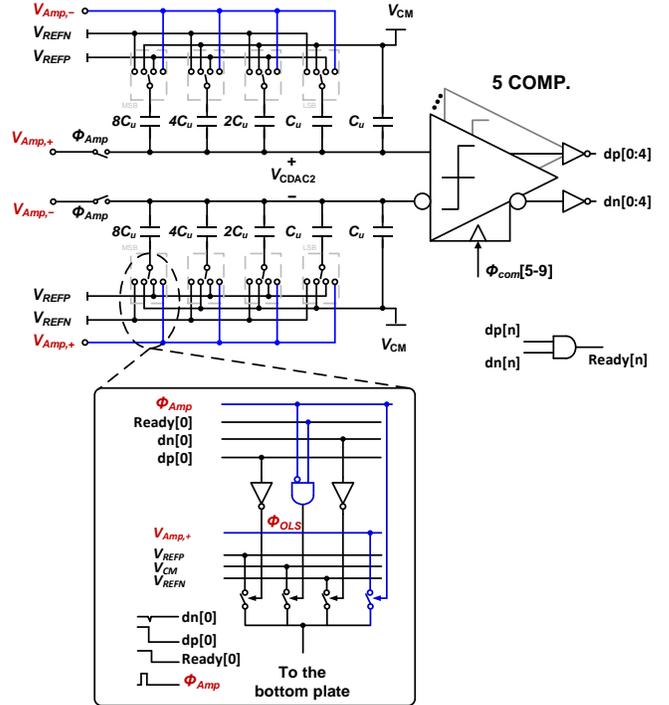

Fig. 6. Second stage reference switch control with embedded OLS technique.

Section II, the second-stage reference voltage values are tuned to accommodate any static deviations in the dynamic amplifier gain due to the exact $\Phi_{Amp}$ pulse width.

### A. Dynamic Amplifier and Comparators

A clocked inverter-based buffer that achieves a gain of ~4 serves as the residue amplifier stage, shown in detail in Fig. 5(a). In addition to the main transconductance transistors M1/2 and M7/8, M3 acts as a current source that is switched on by M4 when $\Phi_{Amp}$ is activated. The gray transistors improve the dynamic performance, with M10-13 compensating kickback noise and M9 boosting the amplifier startup. When the amplifier is disabled transistors M5 and M6 reset $V_{Source}$ to VDD and short the differential output, respectively. As shown in the Fig. 5(b) simulation results, resetting CDAC2 allows the amplifier output to start separating from the common mode and then experience an effective doubling after the level shifting.

Dynamic two-stage comparators are used to allow for low-voltage operation in the two pipeline stages. These comparators are foreground offset-calibrated with current-mode DACs.

### B. Second Stage Reference Switch with Embedded OLS

Fig. 6 illustrates how the OLS technique is included with minor logic changes in the reference switch control to allow both the CDAC2 bottom and top plate to connect to the amplifier output during the amplification phase. As shown in detail for the negative DAC MSB switches, there is an extra right-most switch that connects the bottom plate to the positive input signal when $\Phi_{Amp}$ is high. An AND gate then produces the $\Phi_{OLS}$ signal to switch the bottom plate to the common mode when $\Phi_{Amp}$ goes low and the comparators' ready signals are enabled. Since the extra switch is added to the CDAC bottom plate, there are no speed penalty or reference attenuation issues.

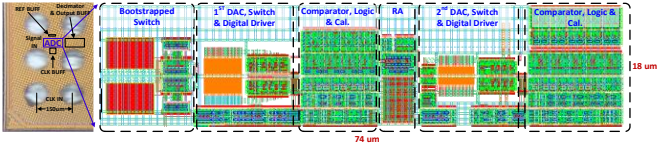

Fig. 7. Prototype ADC chip micrograph.

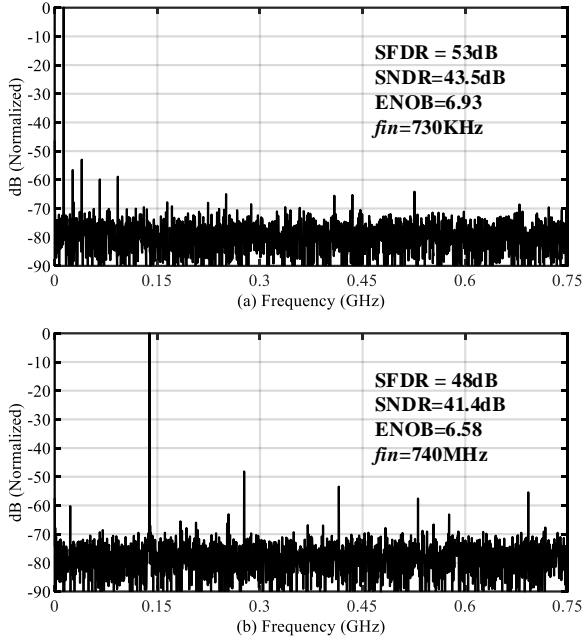

Fig. 8. 1.5GS/s ADC normalized output spectrum for (a) low frequency input and (b) close to Nyquist input. (Decimated by 18, 4096 FFT point).

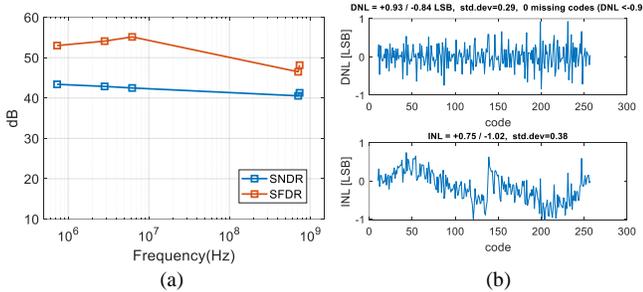

Fig. 9. (a) SNDR and SFDR vs. input frequency. (b) DNL and INL plots.

## IV. EXPERIMENTAL RESULTS

Fig. 7 shows the chip micrograph of the pipelined-SAR ADC, which was fabricated in a 14nm CMOS FinFET process and occupies an active area of 0.0013mm². The ADC is powered from a 0.8V supply and has a 460mV$_{pp,diff}$ full-scale input range with a common mode of 500mV. Testing is performed with an initial foreground offset calibration for all comparators. Fig. 8 shows DFTs of the ADC output when operating at 1.5GS/s. The achieved SNDR is 43.5dB and 41.4dB for low frequency and close to Nyquist inputs, respectively, translating to 6.93 and 6.58 bits ENOB. Fig. 9 shows that the ADC maintains over 40dB SNDR and 45dB SFDR over frequency and that the maximum DNL and INL are +0.93/-0.84 and +0.75/-1.02 LSB, respectively. Table I summarizes the ADC performance and compares this work against previous medium resolution ADCs with GS/s sample rates. The pipelined-SAR architecture with the efficient OLS amplifier allows for 1.5GS/s operation at the lowest 0.8V supply and 2.4mW power consumption, while also achieving the best 16.7 fJ/conv-step FOM.

TABLE I: PERFORMANCE SUMMARY

| References | VLSI'18 [5] | CICC'19 [6] | ISSCC'17 [7] | ISSCC'17 [3] | This Work |
|---|---|---|---|---|---|
| Technology (nm) | 40 | 40 | 28 | 14 | **14** |
| Supply (V) | 1.2 | 1.1 | 0.9 | 0.95 | **0.8** |
| Architecture | Two step SAR | 2-3b SAR | 1-2b SAR | Pipe-SAR | **Pipe-SAR** |
| Channels | 1 | 1 | 2 | 1 | **1** |
| Resolution (bits) | 8 | 7 | 7 | 10 | **8** |
| Sampling rate (GS/s) | 1.1 | 0.9 | 2.4 | 1.5 | **1.5** |
| ENOB @Nyquist | 7.18 | 6.3 | 6.36 | 8 | **6.58** |
| Area (mm²) | 0.00165 | 0.014 | 0.0043 | 0.00158 | **0.0013** |
| Power (mW) | 4 | 2.6 | 5 | 6.9 | **2.4** |
| FOM (fJ/conver-step) | 25 | 36.6 | 25.3 | 17.7 | **16.7** |

## V. CONCLUSION

This paper presented a single channel 8-bit pipelined-SAR ADC that utilizes a novel low-overhead OLS settling technique in the dynamic residue amplifier. A low power design is realized by combining this technique with the use of parallel comparators in the two asynchronous pipeline stages to allow for 1.5GS/s operation with a low 0.8V supply voltage.

ACKNOWLEDGMENT

This work was supported in part by SRC TxACE Grant 2810.013 and NSF Grant 1930828.